Estimation of the partition functions of two-dimensional nearest neighbour Ising models

M V Sangaranarayanan

Department of Chemistry, Indian Institute of Technology- Madras

Chennai-600036 India

E-mail: sangara@iitm.ac.in

**Abstract**

Employing Onsager's exact solution for two- dimensional Ising models, simple expressions are proposed for computing the partition function, magnetization, specific heat and susceptibility for non-zero magnetic fields of square lattices. The partition function in zero fields is also estimated and excellent agreement with the values arising from Onsager's exact solutions is demonstrated.

**1.Introduction**

The extension of Onsager's exact solution of two-dimensional nearest neighbour Ising models for non-zero magnetic field continues to be a frontier area of research in condensed matter physics [1]. The conventional transfer matrix method successfully employed for one-dimensional Ising models is no longer valid for two dimensions [2].

*Prima facie*, it appears that mere formulation of the partition function in itself is inadequate to extract the critical properties in so far as any analytical expressions resulting from the summation of the energies in the Hamiltonian do not possess singularities- essential for the prediction of critical phenomena. However, Onsager's exact solution in this context [3] has not only led to the partition function in zero field but also predicted the critical temperatures exactly.

Here, simple analytical expressions are proposed for the zero-field partition function of two- dimensional Ising models, which yield excellent agreement with Onsager's exact values. The extension of the result to finite magnetic fields is also carried out, thereby yielding the corresponding eqns for the magnetization and susceptibility.

**2. Methodology**

Consider the two-dimensional nearest neighbour Ising Hamiltonian represented as [4]

$$H_T = -J \left[\sum_{\langle ij \rangle} (\sigma_i \sigma_j \sigma_i \sigma_{j+1} + \sigma_i \sigma_j \sigma_{i+1}, \sigma_j)\right] - H \sum \sigma_i \quad (1)$$

where J denotes the interaction energy, H being the external magnetic field. For brevity, a square lattice is considered and the value of Bohr magneton is assumed as 1. The corresponding partition function has been derived as [5]:

$$\left(\frac{1}{N}\right) \ln Q = \ln[\sqrt{2} \cosh(2K)] + (1/\pi) \int_0^{\frac{\pi}{2}} d\phi \ \ln[1 + \sqrt{(1 - \kappa^2 \sin^2(\phi))}] \quad (2)$$

where

$$\kappa = \frac{2 \sinh(2K)}{\cosh^2(2K)} \quad \text{and } K = J/kT \quad (3)$$

**(A) Empirical equations for the partition function**

There exists no simple closed form expression for the integral appearing in eqn (2). However, the numerical evaluation of the integral is straight-forward. Nevertheless, the non-availability of an algebraic expression *masks* the elegance of the partition function equation deduced by Onsager. Furthermore, the integral appearing in eqn (2) is deceptively simple and its evaluation does not appear to be unsurmountable by known methods. By employing a heuristic approach, the integral in eqn (2) has been evaluated and the resulting partition function is given below.

$$\left(\frac{1}{N}\right) \ln Q = \frac{3}{4} \ln 2 + \ln[\cosh(2K)] + (1/4) \ln[1 + \sqrt{(1 - G \kappa^2)}] \quad (4)$$

where $G$ denotes the Catalan constant [6] defined as

$$G = \int_0^{\frac{\pi}{4}} \ln(2 \cos(\phi)) \, d\phi \quad (5)$$

and K and $\kappa$ have been defined in eqn (3).

Table 1 provides the values of the partition function estimated using eqn (4) with Catalan constant as 0.9159. The values computed using eqn (4) are in excellent agreement with those arising from the Onsager's exact solution [3] for values of J/kT ranging from 0 to 2. The corresponding values of κ have also been provided in Table 1. For the sake of clarity, the values have been rounded off to four decimals. The partition functions of two-dimensional Ising models can be represented using Gauss hypergeometric functions [7].

At this stage, it is of interest to re-write eqn (4) as

$$\left(\frac{1}{N}\right)\ln Q = \frac{3}{4}\ln 2 + \ln[\cosh(K)] + \frac{1}{4}\ln\left[1 + \sqrt{\left\{1 - \kappa^2 \int_0^{\frac{\pi}{2}} d\phi \ln[2\cos(\phi)]\right\}}\right] \quad (6)$$

wherein the upper limit of the integral appearing in the Catalan constant (eqn 5) has been changed to π/2 and cosh(2K) is altered to cosh(K). Consequently, the above partition function is identical with that for one dimension at H =0 and J ≠ 0 viz

$$\left(\frac{1}{N}\right)\ln Q = \ln[2\cosh(K)] \quad (7)$$

Thus, the partition function at H =0 for both one- and two-dimensional Ising models viz eqns (5) and (6) have isomorphic structures, when viewed from the perspective offered by eqn (4). Analogously, it is possible to deduce the trivial limit for J =0 and H ≠ 0 by rewriting eqn(6) as

$$\left(\frac{1}{N}\right)\ln Q = \frac{3}{4}[\ln 2] + \ln[\cosh(K)\cosh(H/kT)] + \frac{1}{4}\ln\left[1 + \sqrt{\left\{1 - \kappa^2 \int_0^{\frac{\pi}{2}} d\phi \ln[2\cos(\phi)]\right\}}\right] \quad (8)$$

When K =0, the above reduces to

$$\left(\frac{1}{N}\right)\ln Q = \ln[2\cosh(H/kT)]$$

which is the one-dimensional partition function at J =0 and H ≠ 0.

As mentioned earlier, there are several methods of postulating algebraic equations to estimate the partition functions. The essential pre-requisite in this context (at H =0) are the limiting values at Tc/T = 0 and Tc/T =1. Keeping this in mind, it is possible to postulate the following eqn for the partition function at H =0 for Tc/T ≤ 1.

$$\left(\frac{1}{N}\right)\ln Q = \left\{1 - \frac{T_c}{T}\right\}\ln\left\{1 + \sqrt{(1 - \frac{T_c}{T})}\right\} + f(\sinh(2K)) \quad (9)$$

where f(K) is a polynomial in sinh(2K) given by

f(sinh(2K) = -0.0579857(sinh(2K))$^4$ -0.0411813(sinh(2K))$^3$ -0.0565982(sinh(2K))$^2$
+1.08565(sinh(2K) \hfill (10)

It is easy to note that f(K) = 0 when J/kT = 0 and the first term of eqn (10) yields ln 2, as anticipated. Analogously, when T/Tc =1, the first term of eqn (10) vanishes while the second term yields a value of ≈ 0.9299-in satisfactory agreement with Onsager's exact solution. Table 2 provides the partition functions estimated using eqn (9). It is seen that the agreement with the values arising from Onsager's exact solution is satisfactory.

It is well known that the partition function estimates are not sensitive enough to capture the finer details of the critical phenomena, in general. Nevertheless, eqns (4) and (9) are proposed here with a view to construct suitable Pade' Approximants with desired singularities, whose efficacy are well known in the case of lattice models [8].

A serious limitation of eqns (4) and (9) consists in its inability to predict (i)critical temperature for the magnetization and (ii) logarithmic singularity of the specific heat and(iii) singularity of the susceptibility at H = 0. Thus, it is imperative to propose partition functions which enable the prediction of critical phenomena pertaining to magnetization, susceptibility and specific heat.

**(B) Partition functions for analysing the critical phenomena**

Several approaches are available to derive functions for specific heat that provide logarithmic singularity of the specific heat, the notable among them being the one due to Abe [9]. By following the analysis of Abe [9], a part of the total partition function can be deduced as

$$\frac{\ln Q}{N}(partial)(H=0) = \left(1-\frac{T_c}{T}\right)\ln\left(1-\frac{T_c}{T}\right) + 2\left(\frac{T_c}{T}\right) - Li_2\left(\frac{T_c}{T}\right) + \text{additional terms} \quad (11)$$

where $Li_2$ denotes *dilogarithm* defined as $Li_2 = \int_0^{T_c/T} \frac{\ln(1-t)}{t} dt$

and $T_c/T \leq 1$. The above expression for H = 0 is adequate to deduce the corresponding expressions for the internal energy and specific heat. The internal energy follows from the partition function in the following manner:

$$E = kT^2 \left(\frac{\partial \ln Q}{\partial T}\right)$$

It is easy to verify that eqn (11) predicts the specific heat as

$$C_v/k \sim -\ln\left(1-\frac{T_c}{T}\right) - \frac{T_c}{T} + \text{additional terms involving } \frac{T_c}{T} \quad (12)$$

with a logarithmic singularity. It must be emphasized that the above eqn is only a partial contribution to the specific heat and hence is markedly different from the well-known exact result of Onsager [3].

The total partition function that yields the agreement with Onsager's exact values is as follows:

$$\frac{\ln Q}{N}(H=0) = \left(1-\frac{T_c}{T}\right)\ln\left(1-\frac{T_c}{T}\right) + 2\left(\frac{T_c}{T}\right) - Li_2\left(\frac{T_c}{T}\right) + f(T_c/T) \quad (13)$$

where f(Tc/T) is given by the polynomial

f(Tc/T) = -0.097159441332(Tc/T)$^4$ + 0.09531413951(Tc/T)$^3$ – 0.1267974001(Tc/T)$^2$ + 0.01027566338 (Tc/T) + 0.6931157887     (14)

The partition functions estimated using eqns(13) and (14) are shown in Table 3 and the agreement with Onsager's exact values is excellent. Thus, eqn (13) yields partition functions in satisfactory agreement with exact values apart from the correct prediction of the specific heat behaviour.

### (C)Partition function at H ≠ 0 and spontaneous magnetization

The equation for magnetization can be deduced from the partition function in the thermodynamic limit [5] using

$$M = kT \left(\frac{\partial \ln Q}{\partial H}\right)_T$$

Thus, the partition function at H ≠ 0 may be formulated using Onsager's exact solution for the spontaneous magnetization $M_0$ as well as the exact critical exponents β (1/8) and δ (15) as the input. It is then straight forward to obtain the partition function for H ≠ 0. Thus

$$\frac{\ln Q}{N}(H \neq 0) \approx \frac{M_0 H}{kT} + \left(1 - \frac{T_c}{T}\right)^{7/4} \left[\ln\left[2\cosh\frac{H}{kT(1-\frac{T_c}{T})^{7/4}}\right]\right] - \frac{T_c}{T}\Gamma\left(-\frac{\delta+1}{\delta}, \frac{kT}{H}\right) \quad (15)$$

with Tc/T ≤ 1 and Γ (a, x) denotes the *incomplete Gamma function* [10]. A qualitative justification of the above consists in the limiting case of r h s yielding ln [2cosh(H/kT)] when Tc/T = 0. While the above eqn is applicable for *H > 0*, the earlier eqn (13) is valid for H =0. The magnetization deduced using the above eqn for H >0 follows as

$$M \approx M_0 + \tanh\frac{H}{kT(1-\frac{T_c}{T})^{7/4}} - \frac{T_c}{T}e^{-kT/H}\left(\frac{H}{kT}\right)^{\frac{1}{\delta}} \quad (16)$$

where the value of 7/4 incorporates the critical exponents β and δ denote the respective values 1/8 and 15 arising from Onsager's exact solution. It is easy to verify that at H =0, the above eqn yields the spontaneous magnetization $M_0$ and the exact eqn for the same has been proposed by Onsager [3] for square lattices:

$$M_0 = \{1 - \sinh^{-4}(2K)\}^{1/8}$$

Furthermore, $M_0$ = 0 when T/Tc ≥ 1 and hence eqn (16) predicts magnetization values <1. Analogously, at T = T$_c$, the magnetization yields the limit as

$$M \sim H^{1/\delta}$$

with δ denoting the critical exponent for magnetization (δ =15 in the case of the exact analysis of two-dimensional nearest neighbour Ising models).

Table 4 provides the value of the magnetization, estimated using eqn(16), for $T_c/T = 0.4$. The field-dependent susceptibility is easily obtained as

$$kT \chi \approx \left\{ \frac{1}{(1-\frac{T_c}{T})^{7/4}} \right\} sech^2 \frac{H}{kT(1-\frac{T_c}{T})^{7/4}} - \frac{T_c}{T} \left\{ e^{-\frac{kT}{H}} \left[\frac{H}{kT}\right]^{(\frac{1}{\delta}-1)} \right\} \left\{ \frac{1}{\delta} + \frac{kT}{H} \right\} \quad (17)$$

The zero-field susceptibility follows as

$$\chi(H=0) \sim \frac{1}{(1-\frac{T}{T_c})^{\beta(\delta-1)}}$$

with the value of the critical exponent γ being 7/4. Table 5 summarizes the expressions for various quantities within the two-dimensional nearest neighbour Ising models for square lattices, emphasizing the approximate validity of these being restricted to $T_c/T \leq 1$ and H>0.

The foregoing analysis has provided simple algebraic expressions for the partition functions, specific heat, magnetization and susceptibility for two-dimensional nearest neighbour Ising models. The computed partition functions for zero magnetic fields are in excellent agreement with those arising from Onsager's exact solution.

4. Summary

The partition functions of two-dimensional nearest neighbour Ising models at H =0 are proposed and excellent agreement with Onsager's exact solution is inferred. The field-dependent magnetization and susceptibility have been derived, yielding the correct critical exponents.

Table 1-Estimation of the partition function using eqn (4) and comparison with Onsager's exact solution

| J/kT | $\kappa$ | From eqn (4) | Onsager's exact solution |
| --- | --- | --- | --- |
| 0 | 0 | 0.6931 | 0.6931 |
| 0.1 | 0.3869 | 0.7039 | 0.7032 |
| 0.2 | 0.7029 | 0.7362 | 0.7345 |
| 0.3 | 0.9060 | 0.7910 | 0.7906 |
| 0.4 | 0.9930 | 0.8784 | 0.8794 |
| 0.5 | 0.9871 | 1.0246 | 1.0258 |
| 0.6 | 0.9208 | 1.2100 | 1.2101 |
| 0.7 | 0.8232 | 1.4057 | 1.4042 |
| 0.8 | 0.7151 | 1.6034 | 1.6018 |
| 0.9 | 0.6093 | 1.8023 | 1.8008 |
| 1.0 | 0.5125 | 2.0015 | 2.0003 |
| 1.1 | 0.4272 | 2.2010 | 2.2001 |
| 1.2 | 0.3540 | 2.4006 | 2.4001 |
| 1.3 | 0.2922 | 2.6004 | 2.6000 |
| 1.4 | 0.2405 | 2.8002 | 2.8000 |
| 1.5 | 0.1976 | 3.0002 | 3.0000 |
| 1.6 | 0.1622 | 3.2000 | 3.2000 |
| 1.7 | 0.1330 | 3.4000 | 3.4000 |
| 1.8 | 0.1090 | 3.6000 | 3.6000 |
| 1.9 | 0.0893 | 3.8000 | 3.8000 |
| 2.0 | 0.0731 | 4.0000 | 4.0000 |

Table 2 – Comparison of the partition functions estimated using eqn(9) with the values arising from Onsager's exact solution for (Tc/T) ≤ 1

| $T_c/T$ | Values of $\left(\frac{1}{N}\right)\ln Q$ from eqn(9) | Values $\left(\frac{1}{N}\right)\ln Q$ from Onsager's exact solution |
|---|---|---|
| 0 | 0.69314 | 0.69314 |
| 0.2269 | 0.70354 | 0.70323 |
| 0.3022 | 0.73410 | 0.73453 |
| 0.68076 | 0.79107 | 0.79056 |
| 0.90768 | 0.87910 | 0.87936 |
| 1 | 0.92988 | 0.92969 |

Table 3 – Comparison of the partition functions estimated using eqn(13) with the values arising from Onsager's exact solution for (Tc/T) ≤ 1

| $T_c/T$ | Values of $\left(\frac{1}{N}\right)\ln Q$ from eqn(13) | Values $\left(\frac{1}{N}\right)\ln Q$ from Onsager's exact solution |
|---|---|---|
| 0 | 0.69314 | 0.69314 |
| 0.2269 | 0.70335 | 0.70323 |
| 0.4538 | 0.73426 | 0.73453 |
| 0.68076 | 0.79086 | 0.79056 |
| 0.90768 | 0.87909 | 0.87936 |
| 1 | 0.92969 | 0.92969 |

Table 4- Magnetisation values estimated using eqn for Tc/T = 0.4 and various values of H/kT

| Dimensionless magnetic field (H/kT) | Magnetization |
|---|---|
| 0 | 0 |
| 0.2 | 0.159 |
| 0.3 | 0.351 |
| 0.4 | 0.499 |
| 0.5 | 0.608 |
| 0.6 | 0.687 |
| 0.7 | 0.743 |
| 0.8 | 0.784 |
| 0.9 | 0.814 |
| 1.0 | 0.838 |

Table 5 – Equations for the partition function, magnetization and susceptibility for square lattices in the two-dimensional Ising model

| Parameter | Governing Equation |
|---|---|
| partition function for H ≠ 0 | $\frac{\ln Q}{N} \approx \frac{M_0 H}{kT} + (1 - \frac{T_c}{T})^{7/4} \left[ \ln \left[ 2 \cosh \frac{H}{kT(1-\frac{T_c}{T})^{7/4}} \right] \right] - \frac{T_c}{T} \Gamma(-\frac{\delta+1}{\delta}, \frac{kT}{H})$ |
| partition function for H =0 | $\frac{\ln Q}{N}(H=0) = \left(1 - \frac{T_c}{T}\right) \ln \left(1 - \frac{T_c}{T}\right) + 2\left(\frac{T_c}{T}\right) - \text{Li}_2\left(\frac{T_c}{T}\right) + f(T_c/T)$ |
| magnetization at H ≠ 0 | $M \approx M_0 + \tanh \frac{H}{kT(1-\frac{T_c}{T})^{7/4}} - \frac{T_c}{T} e^{-kT/H} \left(\frac{H}{kT}\right)^{\frac{1}{\delta}}$ |
| magnetic susceptibility at H ≠ 0 | $kT\chi \approx \left\{ \frac{1}{(1-\frac{T_c}{T})^{7/4}} \right\} \text{sech}^2 \frac{H}{kT(1-\frac{T_c}{T})^{7/4}} - \frac{T_c}{T} \left\{ e^{-\frac{kT}{H}} \left[\frac{H}{kT}\right]^{(\frac{1}{\delta}-1)} \right\} \left\{ \frac{1}{\delta} + \frac{kT}{H} \right\}$ |

Acknowledgements

This work was supported by MATRICS scheme of Science and Engineering Research Board, Government of India.